\documentclass[twocolumn,prb]{revtex4}
\usepackage{epsfig}
\usepackage{graphicx}
\usepackage{amsmath}
\usepackage{amssymb}

\begin{document}


\title{From Discrete Hopping to Continuum Modeling on Vicinal Surfaces with
Applications to Si(001) Electromigration}

\author{Tong Zhao$^1$}
\author{John D. Weeks$^2$}
\affiliation{$^{1,2}$Institute for Physical Science and Technology, University
of Maryland, College Park, Maryland 20742}
\address{$^2$Department of Chemistry and Biochemistry,
University of Maryland, College Park, Maryland 20742}
\author{Daniel Kandel$^3$}
\affiliation{$^3$Department of Physics of Complex System,
Weizmann Institute of Science,
Rehovot 76100, Israel}
\altaffiliation[Present and permanent address: ]{KLA-Tencor Corporation,
Hatikshoret St, Migdal Haemek, Israel}
\date{\today }

\begin{abstract}
Coarse-grained modeling of dynamics on vicinal surfaces concentrates on the
diffusion of adatoms on terraces with boundary conditions at sharp
steps, as first studied by Burton, Cabrera and Frank (BCF). Recent
electromigration experiments on vicinal Si surfaces suggest the need for
more general boundary conditions in a BCF approach. We study a discrete
1D hopping model that takes into account asymmetry in the hopping rates
in the region around a step and the finite probability of incorporation
into the solid at the step site. By expanding the continuous concentration
field in a Taylor series evaluated at discrete sites near the step,
we relate the kinetic coefficients and
permeability rate in general sharp step models to the physically
suggestive parameters of the hopping models. In particular we find that
both the kinetic coefficients and permeability rate can be negative when
diffusion is faster near the step than on terraces.
These ideas are used to provide an understanding of recent electromigration
experiment on Si(001) surfaces where step bunching is induced by an electric 
field directed at various angles to the steps.
\end{abstract}

\maketitle

\section{Introduction}

Steps on vicinal crystal surfaces, created by a miscut along a low index
plane, have long been of great interest in both basic and applied research.%
\cite{vicinalsurfacereview} High quality crystals can be grown through
step-flow --- the uniform motion of more or less equally-spaced steps ---
and step bunching instabilities can create arrays of wide flat terrace
separated by closely bunched steps. Other arrangements of steps could serve
as templates for nanoscale structures and devices.

Most fundamental studies of the static and dynamic properties of vicinal
surfaces are based on generalizations of the classic theory of Burton,
Cabrera, and Frank (BCF), developed more than fifty years ago. \cite
{BCF_classic} This theory describes the diffusion of adatoms on terraces
with boundary conditions at steps, which are treated as sharp line
boundaries. Originally BCF assumed that the steps acted as perfect sinks and
sources of adatoms so that the limiting adatom concentration at the step
boundaries always reduces to local equilibrium.

Many extensions and modifications of the BCF theory have been suggested to
provide a more general framework for the description of different
experiments. One of the most important was Chernov's introduction of linear
{\em kinetic coefficients},  which permit deviations from local equilibrium at
steps. \cite{k_Chernov, k_Chernov2} It was soon recognized that in general
the kinetic coefficients could be asymmetric.\cite{ESclassic_Schwoebel}
Another generalization permits step {\em permeability}
or {\em transparency}, with a 
term in the boundary condition directly connecting the limiting adatom 
concentration on adjacent terraces.\cite{Permeabilityfirst_Ozdemir} 
These generalized BCF models provide a mesoscopic or coarse-grained
description of surface evolution with effective boundary conditions at sharp
steps, and we will generally refer to them as sharp step models.

Many kinetic instabilities seen in experiments have been successfully
described from this perspective using various combinations of boundary
conditions. However in general it is not clear how to connect the choices and
values of the effective parameters in sharp step models to the underlying
physical processes or how to determine the uniqueness of such a mapping. A
similar difficulty arises in trying to relate ``microscopic'' parameters in
kinetic Monte Carlo simulations of discrete hopping models
to the effective parameters in a generalized
sharp step model. Very different microscopic models can sometimes seem to
give equally plausible mesoscopic descriptions of limited sets of
experimental data.

In previous work \cite{TZ_em_short} we proposed a novel continuum two-region
diffusion model (CTRM), which gave a rather simple and unified description
of a variety of current-induced instabilities seen experimentally on vicinal
Si surfaces. We will discuss these experiments in more detail later. The
model assumes that diffusion rates in a finite region around a step could be
affected by the different local bonding configurations and thus differ from
those found elsewhere on terraces. By extrapolating the steady state
concentration profiles to the center of the step region, we obtained a
mapping of the parameters in the CTRM to those of an equivalent classical
sharp step model. One surprising conclusion was that {\em negative} kinetic
coefficient can arise when the diffusion rate near a step is faster than
that on the terraces.

In this paper, we will provide a more systematic way of deriving the
boundary conditions for the continuum sharp step models from a rather
general 1D discrete hopping model that permits both asymmetric diffusion in
the step region as well as step permeability. As discussed by Ghez and
Iyer,\cite{MBEgrowth_Ghez} such an effective 1D model can result from 
averaging over relevant 2D configurations of kink and ledge sites on an 
atomic step. We then use these ideas to provide a detailed description of 
electromigration on Si(001) surfaces, and find a coherent scenario that 
explains most of the interesting experimental findings.

\section{1D Hopping Model} \label{sec:hopping_generic}

\begin{figure}[tbp]
\includegraphics[width=76mm,height=57mm]{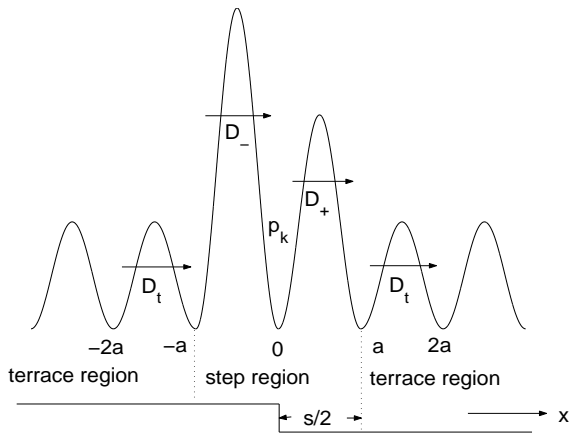}
\caption{A schematic plot of the 1D potential surface near an atomic step.
Different $D$'s that have dimensions of diffusion constants
characterize the hopping rates associated with different barrier
heights. Here, we take the width of the step
region to be $2a$.}
\label{Fig1nonps_generic}
\end{figure}

The simple 1D model that we study is schematically shown in Fig. \ref
{Fig1nonps_generic}, where an atomic step site is surrounded by a region of
width $s$ with generally different diffusion rates, induced by
reconstruction or rearrangements of local dangling bonds. As we will see,
this difference can generate effective kinetic coefficients in a sharp step
description.

In general, the width $s$ of the step region with different diffusion
barriers should vary for different systems. However, we find that the
essential physics of the hopping model is not strongly affected by specific
choices of $s$ of order of a few lattice spacings $a$. Thus we analyze the
algebraically simplest case shown in Fig. \ref{Fig1nonps_generic} with
half-step regions of width $a$. The more general result, needed in the
analysis of Si(001) experiments, is easily obtained by replacing $a$
with $s/2$, as will become clear later when we compare the results of this
generic hopping model to our previous results.

We include here two additional physical features of the step region as
illustrated in Fig. \ref{Fig1nonps_generic}. One is the possible asymmetry in
the diffusion processes in the up and down half-step regions, described by
hopping rates $D_{\pm }/a.$ The $D_{\pm }$ have dimensions of a diffusion
constant, and the model is usefully characterized by dimensionless
parameters 
\begin{equation}
R_{\pm }\equiv D_{\pm }/D_{t},  \label{Rplusminus}
\end{equation}
with $D_{t}$ the diffusion constant on the terraces.

The other feature we build in is {\em step permeability} or {\em transparency},
characterized in our model by a single parameter $p_{k}$ ($0\leq
p_{k}\leq 1$). This can be understood as the effective probability in our 1D
model that an adatom hopping to site $0$ will encounter a kink site at a
given temperature and thus equilibrate with the solid. This parameter takes
account of effects from both kink site density and ledge diffusion in a full
2D model. When $p_{k}=1$, the step site acts as a perfect sink maintained by
either enough kink sites or fast ledge diffusion or both, and consequently
the step site concentration will be pinned at equilibrium $c_{eq}$. In the
opposite limit with $p_{k}=0$ no adatoms are incorporated into the solid.
The step site behaves like any other terrace site and thus is perfectly
permeable. We neglect other possible sources of permeability, including
direct hopping over the step region from one terrace to another or effects
of rapid step motion, \cite{Phasefieldsteps_Pierre}
which we believe are less physically relevant for our cases of interest.
In this section we will also consider only diffusion fluxes from concentration
gradients, and discuss the case of driven diffusion from an external
field in the Appendix.

We assume that the net flux of adatoms that hop between step site $0$ and
site $a$ can be partitioned into two effective contributions: 
\begin{equation}
J_{a/2}=\frac{D_{+}}{a}\left[ p_{k}\{c_{eq}-\hat{c}\left( a\right) \}+\left(
1-p_{k}\right) \{\hat{c}\left( 0\right) -\hat{c}\left( a\right) \}\right] .
\label{eq:Ja/2}
\end{equation}
The first term describes an adatom exchange with probability $p_{k}$
involving equilibrated ``kink-like'' adatoms at site $0$ with density $%
c_{eq} $ and the neighboring terrace site. The second term involves a
similar exchange with probability $\left( 1-p_{k}\right) $ involving
unincorporated ``ledge-like'' adatoms with density $\hat{c}\left( 0\right) .$
Only the former involves creation/annihilation of adatoms, and the latter is
treated as a normal diffusion flux that conserves the adatom density.

Similarly, the flux from site $-a$ to $0$ is 
\begin{equation}
J_{-a/2}=\frac{D_{-}}{a}\left[ p_{k}\{\hat{c}\left( -a\right)
-c_{eq}\}+\left( 1-p_{k}\right) \{\hat{c}\left( -a\right) -\hat{c}\left(
0\right) \}\right] .  \label{eq:J-a/2}
\end{equation}
Since we assume that all the sinks/sources reside only at site $0$, the net
flux of adatoms that hop from site $a$ ($-2a$) to site $2a$ ($-a$) takes on
the simpler form 
\begin{equation}
J_{\pm 3a/2}=\pm \frac{D_{t}}{a}\left[ \hat{c}\left( \pm a\right) -\hat{c}
\left( \pm
2a\right) \right] .  \label{eq:J3a/2}
\end{equation}

As in many other situations of physical interest, we will use the {\em %
quasi-static }approximation to simplify the analysis. Here we assume that the
motion of the step region is much slower than the relaxation of the terrace
diffusion field, so that one can determine the diffusion process on terrace
sites with fixed positions of the step regions. In the quasi-static limit the
net change in the number of adatoms at each terrace site given by a total
flux balance must vanish, i.e., $d\hat{c}(x)/dt=0$ for all $x=\pm a,\pm
2a\dots $ In particular, at sites $\pm a$, the balance of fluxes is given by 
\begin{equation}
0=\frac{d\hat{c}\left( \pm a\right) }{dt}=\pm \left[ aJ_{\pm a/2}-aJ_{\pm
3a/2}\right] .  \label{massblna}
\end{equation}

At step site $0$, $\hat{c}\left( 0\right) $ can be determined by balancing
the conserved flux terms proportional to $\left( 1-p_{k}\right) $ in Eqs. (%
\ref{eq:Ja/2}) and (\ref{eq:J-a/2}), and is given by 
\begin{equation}
\hat{c}\left( 0\right) =\frac{D_{+}\hat{c}\left( a\right) +D_{-}\hat{c}%
\left( -a\right) }{D_{+}+D_{-}}.  \label{eq:c0}
\end{equation}

\section{Relating parameters in discrete and continuum models} \label{sec:discretecont}

Our task now is to relate the physically suggestive parameters $R_{\pm }$
and $p_{k}$ in the discrete hopping model to the kinetic coefficients $%
k_{\pm }$ and permeability rate $P$ appearing in the boundary conditions of
a continuum sharp step model as in Eq. (\ref{eq:BCFgen}) below. For $x>0$,
consider a smooth continuum concentration profile $c\left( x\right) $ that
passes through the discrete concentrations $\hat{c}\left( a\right) $ and
$\hat{c}\left( 2a\right) $. (The caret distinguishes discrete from continuum
functions.) The behavior of $c\left( x\right) $ at larger $x$ is determined
by the physical processes on the terraces, 
but does not need to be specified explicitly for our purposes here.

To make contact with the sharp step model, we rewrite the fluxes in
Eqs. (\ref{eq:Ja/2})-(\ref{massblna}) in terms of $c(x)$.
To that end we use a Taylor series expansion
to linear order to express $c\left( a\right) =\hat{c}\left( a\right) $ and
$c\left( 2a\right) =\hat{c}\left( 2a\right) $ in terms of $c^{+}\equiv
c(0^{+}),$ the extrapolated limiting concentration as $x\rightarrow 0^{+}$
at the sharp step edge in a continuum picture, and its associated gradient
$\nabla c\mid _{+}$. Similarly, $\hat{c}\left( -a\right) $ and $\hat{c}\left(
-2a\right) $ can be expressed in terms of $c^{-}$ and $\nabla c\mid _{-}$,
which in general are different than $c^{+}$ and $\nabla c\mid _{+}$.

Using Eq. (\ref{eq:c0}) to eliminate $\hat{c}\left( 0\right) ,$ and
substituting into Eq. (\ref{massblna}), we find that the result can be
rewritten in the form of a generalized {\em linear kinetics boundary condition}
\cite{MBEgrowth_Ghez} with permeability 
\begin{equation}
\pm \left[ D_{t}\nabla c\mid _{\pm }\mp vc^{\pm }\right] =k_{\pm }\left( c^{%
\pm }-c_{eq}\right) +P\left( c^{\pm }-c^{\mp }\right) .  \label{eq:BCFgen}
\end{equation}
The kinetic coefficients $k_{\pm }$ are given by 
\begin{equation}
k_{\pm }=\frac{D_{t}}{a}\frac{p_{k}}{\left( R_{\pm }-1\right) \left[
1+\left( 1-p_{k}\right) M\right] },  \label{eq:k+}
\end{equation}
where
\begin{equation}
M\equiv \frac{R_{+}R_{-}}{\left( R_{+}+R_{-}\right) }\left[ \frac{R_{+}}{%
R_{-}-1}+\frac{R_{-}}{R_{+}-1}\right]   \label{eq:M}
\end{equation}
is symmetric on exchange of $+$ and $-.$ Note in general that the ratio of
the kinetic coefficients satisfies
\begin{equation}
\frac{k_{+}}{k_{-}}=\frac{R_{-}-1}{R_{+}-1}  \label{eq:k+/k-}
\end{equation}
independent of $p_{k}.$ The permeability rate $P$ can be written as

\begin{equation}
P=\frac{k_{\pm }}{p_{k}(R_{\mp }-1)}\cdot \frac{\left( 1-p_{k}\right)
R_{+}R_{-}}{\left( R_{+}+R_{-}\right) }.  \label{eq:P}
\end{equation}
Using Eq. (\ref{eq:k+}) in the first factor, we see that $P$ is {\em %
symmetric} on exchange of $+$ and $-$, and has a finite limit as $%
p_{k}\rightarrow 0.$

The final parameter $v$ in Eq. (\ref{eq:BCFgen}) is
zero in our present treatment since we used the quasi-static 
approximation to derive Eqs. (\ref{massblna}) and (\ref{eq:c0}). 
In principle, a non-vanishing $v$ would arise if we took the flux due to 
step motion into account in the discrete hopping model. However, the 
quasi-static limit is valid in most physical cases of interest, and thus 
this additional complication can be avoided.

Equations (\ref{eq:BCFgen}-\ref{eq:P}) are the central results in this
section. As mentioned earlier, we find that the sharp step boundary
condition can indeed be generally expressed using linear kinetics with
permeability. More importantly, we
are able to relate the effective parameters in the sharp step boundary
conditions to the physically suggestive parameters we considered in our
generic hopping model. This mapping provides a simple way to understand many
aspects of electromigration phenomena on Si surfaces.

A notable general feature of these equations is that the kinetic
coefficients $k_{\pm }$ are proportional to $p_{k}$ and the permeability
rate $P$ is proportional to $\left( 1-p_{k}\right) $.
The kinetic coefficients characterize adatom exchange
involving equilibrated solid adatoms at kinks and the adatom gas phase,
while the permeability rate characterizes adatom motion across the step
without equilibrating with the solid phase. Moreover, the kinetic
coefficients $k_{\pm }$ are in general asymmetric on the two sides of the
step due to the asymmetry of emission and diffusion processes from kinks.
However, the permeability rate $P$ is symmetric since the physical processes
of hopping from one side to the other without attachment at the step always
involves the diffusion constants on both sides.

We now consider some limits of the above general expressions to illustrate
some interesting features of both the kinetic coefficients and the
permeability rate.

\subsection{Impermeable steps, $p_{k}\rightarrow 1$} \label{sec:ps}

This limit is
usually considered in treatments of the sharp step model, and we used it
to analyze current-induced instabilities on Si surfaces.
\cite{ TZ_em_short} In this limit
the only way for the adatoms to go across a step is through
attachment/detachment at kinks, and the permeability rate $P$ vanishes.

The results are conveniently described in terms of the
{\em attachment/detachment lengths} 
\begin{equation}
d_{\pm }\equiv D_{t}/k_{\pm }\,.  \label{attachdetachlength}
\end{equation}
Using Eq. (\ref{eq:k+}), these are given by 
\begin{equation}
d_{\pm }=a\left( R_{\pm }-1\right) .  \label{dps}
\end{equation}
If we replace the width $2a$ of the step region in the present model by
a general value $s$, we recover exactly the results we found earlier
\cite{TZ_em_short} using the CTRM. As shown in the appendix, Eq. (\ref{dps})
also holds to lowest order even when there is an external driving field
that affects processes in the step region.

This shows the general validity of the mapping between model
parameters that we found earlier by considering diffusion driven by a weak
electric field. For $R_{\pm }>1$, corresponding to slower diffusion in the
step region, the attachment/detachment lengths and kinetic coefficients are
positive. The kinetics is usually called {\em attachment/detachment limited} when
$d_{\pm }$ $\gg l$, with $l$ the average terrace width in a uniform step
train, or {\em diffusion limited} when $0\leq d_{\pm }\ll l$. For $R_{\pm }=1$,
$d_{\pm }$ vanishes, and the kinetic coefficients diverge. This forces $c^{%
\pm }$ to equal $c_{eq}$ in Eq. (\ref{eq:BCFgen}) and generates the local
equilibrium boundary condition originally proposed in the BCF model.

More interestingly, for $R_{\pm }<1$, corresponding to faster diffusion in
the step region, the attachment/detachment lengths and the corresponding
kinetic coefficients are {\em negative}. As we showed earlier,
\cite{TZ_em_short} the sign of
the kinetic coefficients plays a key role in interpreting electromigration
experiments on Si surfaces, since it determines the stability of a uniform
step train for a given current direction. This application of our general
results will be discussed in more detail below.

In the following, we will characterize the limit $p_{k}\rightarrow 1$ as
defining a {\em perfect sink} model, since adatoms can not diffuse across a
step without attachment/detachment at kink sites. As a direct consequence,
the two sides of the step are decoupled and any change of the microscopic
rates on one side of the step does not affect the kinetic coefficient on the
other side. However, as shown
above, the two sides of the step will in general be coupled for $p_{k}<1$
through Eqs. (\ref{eq:k+/k-}) and (\ref{eq:P}), and the subsequent analysis
of step dynamics becomes much more involved.

\subsection{Very permeable steps, $p_{k}\rightarrow 0$}

This limit may be physically relevant at low enough temperatures, or slow
enough ledge diffusion, or some proper combination of both. Here the adatoms
hop around on the surface without encountering sinks/sources in the step
region. Thus one expects vanishing kinetic coefficients, but a finite
permeability rate, and this is indeed what Eq. (\ref{eq:k+}) and Eq. (\ref
{eq:P}) predict in this limit.

As in Eq. (\ref{attachdetachlength}), let us define a corresponding
{\em permeability length} 
\begin{equation}
d_{P}\equiv D_{t}/P.  \label{permeabilitylengthdef}
\end{equation}
Then Eqs. (\ref{eq:P}) and (\ref{eq:k+}) yield 
\begin{equation}
d_{P}=2a\left[ \frac{1}{2}\left( R_{+}+R_{-}\right) -1\right] .
\label{permeabilitylengthR}
\end{equation}
Similar to $d_{\pm },$ the permeability length $d_{P}$ can become {\em %
negative} when $\left( R_{+}+R_{-}\right) <2$, with faster diffusion in the
step region. Eq. (\ref{permeabilitylengthR}) is consistent with results
derived from a continuum phase field model.
\cite{Phasefieldsteps_Pierre} Recently Pierre-Louis and M\'{e}tois
\cite{Correspondence_Pierre} have argued
that negative permeability lengths can explain some novel growth-induced 
instabilities seen during electromigration on Si(111) surfaces.

\begin{figure}[tbp]
\includegraphics[width=76mm,height=50mm]{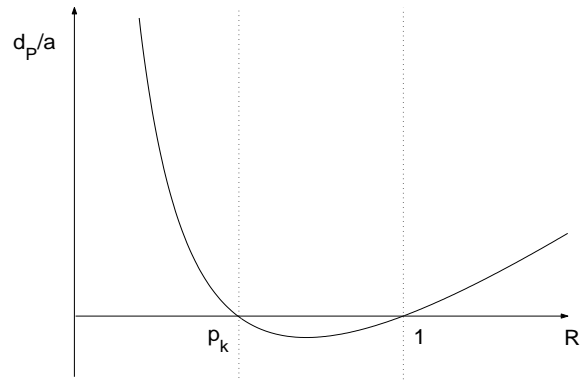}
\caption{Plot of the dimensionless permeability length $d_{P}/a$ as a
function of $R$ in the symmetric case for a general $p_k$.}
\label{Fig2p_alpha}
\end{figure}

\subsection{Partially permeable steps, $0<p_{k}<1$}

This is the most general case, where only a finite fraction of adatoms at
the step equilibrate at kinks, presumably corresponding to intermediate
temperatures with moderate ledge diffusion. We focus on the simplest {\em %
symmetric case} where $D_{\pm }=D_{s}$ or $R_{+}=R_{-}=R$ in Eqs. (\ref{eq:k+}%
)-(\ref{eq:P}). The attachment/detachment length becomes 
\begin{equation}
d=\frac{a}{p_{k}}\left( R-p_{k}\right) ,  \label{eq:k}
\end{equation}
and the permeability length is 
\begin{equation}
d_{P}=2a\left( R-1\right) \frac{\left( R-p_{k}\right) }{\left(
1-p_{k}\right) R}.  \label{eq:Pgen}
\end{equation}

Equation (\ref{eq:k}) can be understood using the same physics as in the
perfect sink model. With a finite probability $p_{k}$ to encounter a kink,
an adatom has to move faster in the step region ($D_{s}=D_{t}/p_{k}$) to
maintain local equilibrium ($d\rightarrow 0$ or $k\rightarrow \infty $)
compared with the perfect sink case ($D_{t}=D_{s}$). The permeability length
in Eq. (\ref{eq:Pgen}) is a new feature arising from the possibility that
the adatoms go directly across the step without equilibrating with the
solid. This expression shows a fairly complicated dependence on microscopic
motions characterized by $R$ and $p_{k}$.

A schematic plot of $d_{P}$ versus $R$ for a given $p_{k}$ is shown in Fig. 
\ref{Fig2p_alpha}. Both $d$ and $d_{P}$ diverge as $R\rightarrow \infty $,
since all motion in the step region vanishes in this limit. $d_{P}$ decreases
as $R$ decreases, and stays positive for $R > 1$. Just like the
attachment/detachment length, the permeability length changes sign from
positive to negative as $R$ passes through $1$, with equal hopping rates in
the terrace and step regions. However, the permeability length becomes
positive again for small enough $R$ when the motion in the step region
is sufficiently fast ($R<p_{k}$) that the probability of crossing the
step without involvement of a
kink is effectively decreased to a point that it is no longer faster than
hopping on terraces.

\section{Applications to Electromigration on Si(001)}\label{sec:emSi001}

We now apply these ideas to
current-induced instabilities on vicinal Si surfaces.
\cite{EMreview_Yagi} Step bunching is seen on Si(111) surfaces when the 
electric current is properly directed normal to the step direction.
\cite{Latyshev89} The uniform step train is initially stable when the 
current flows in the opposite direction. There are
three temperature ranges between about $850^{\circ }C$ and $1300^{\circ }C$
where the stable and unstable directions are reversed.
However, at similar temperatures vicinal 
Si(001) surfaces miscut along $\left[ 110\right] $ exhibit step bunching 
from current normal to the steps in {\em both} directions.
\cite{Si001bchexp_Doi,Si001bchexp_Latyshev}
Characteristic bunching patterns have
also been observed for current directed at various angles to the steps.
\cite{Si001bchexp_Nielsen}  

There is general agreement that in the presence of an electric field
adatoms acquire an effective charge $z^{*}e$ 
(which includes both electrostatic and a ``wind-force'' contribution arising 
from scattering of charge carriers), and thus experience a field-directed 
force ${\bf F}=z^{*}e {\bf E}$ that biases their diffusive motion. However, 
it is less clear what are appropriate boundary conditions
in a sharp step model for these processes and how they might be affected
by the electric field and by surface reconstruction. 
The generic hopping model studied in Sec. 
\ref{sec:hopping_generic} helps us shed some light on these issues,
and the results can be applied to 
electromigration on both Si(111) and Si(001) surfaces. In this paper we 
discuss applications to Si(001) surfaces. The different Si(111) 
instabilities will be discussed elsewhere. \cite{em_long}  

\begin{figure}[tbp]
\includegraphics[width=76mm,height=60mm]{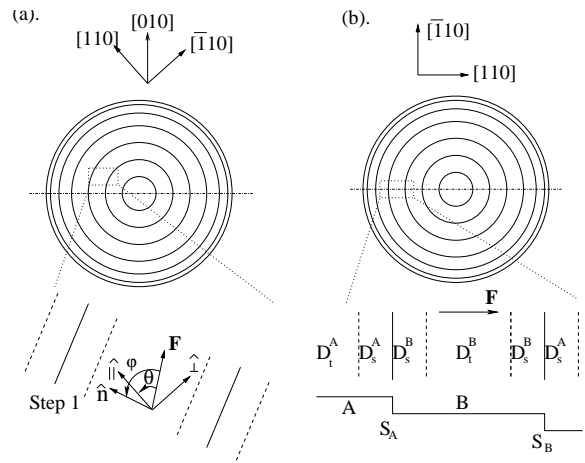}
\caption{A schematic illustration of the dimple
geometry on the Si(001) surface. 
(a) The general view of the dimple with the crystallographic directions 
indicated above. Zooming into a given local area of the dimple (the dotted 
line box), we show the step-terrace configuration with a general direction 
of the electric field. $\varphi$ is the angle between field direction and the
local normal to the steps, while $\theta$ is the angle between the field 
and [110]. $\theta=\pi/4$ corresponds to a field 
direction along $\left[ 010\right]$.
(b) The top view of the dimple when $\theta=0$. Zooming into the 
dotted-lined box near the center of the dimple with $ \varphi=0$, 
we show a top view of the vicinal surface and a side view of the 
step-terrace configuration. Most of basic physics of step pairing and 
bunching will be illustrated in this simple 1D geometry with the electric 
field perpendicular to average step position.}
\label{Fig3schm_dimple2}
\end{figure}

The most notable differences in current-induced step bunching on Si(001)
and Si(111) surfaces arise from the $\left( 2\times 1\right) $
surface reconstruction (dimerization) on Si(001), which persists up to
temperatures of at least $1200^{\circ }C$.\cite{Si001energetics_rev}
Two characteristic directions on the surface are
established by dimerization, either parallel or perpendicular to the
substrate dimer rows in the orthogonal $\left[ 110\right] $ direction, denoted
by $\parallel $ and $\perp $ respectively. Experimental evidence suggests
that the diffusion along the dimer rows is much faster at low temperatures,
i.e., $D_{t}^{\parallel }\gg D_{t}^{\perp }$.\cite{Si001diffanisotropyexp_Mo}

In recent experiments, \cite{Si001bchexp_Nielsen} the bunching behavior
was studied on dimple geometries, where steps of all orientations are found. As
schematically shown in Fig. \ref{Fig3schm_dimple2}a, there are in general two 
angles needed to describe the local geometry of the dimple when the electric field 
is applied,\cite{Si001bchexp_Nielsen} characterized
by the angle $\theta$ between direction of
the electric field and the [110] direction, and the angle $\varphi$ between the
field and the local normal to the steps.

The bunching exhibits interesting angular dependences. When the current is 
parallel to the orthogonal $\left[ 110\right]$ direction ($\theta=0$), 
the bunching is observed to be strongest in the areas where the current is locally 
parallel to the step normal direction ($\varphi=0$), e.g., the dotted line box 
in Fig. \ref{Fig3schm_dimple2}b. No bunching occurs in the corresponding 
perpendicular directions ($\varphi=\pi/2$). However, if the current is 
rotated to $\pi/4$ off the dimer row direction ($\theta=\pi/4$), the 
strongest bunching occurs in the areas where the current is perpendicular 
to the local step directions ($\varphi=\pi/2$). No bunching is seen
in the corresponding perpendicular direction ($\varphi=0$), which in
the previous case was where the maximum bunching was found.
In the following discussion, we will first 
study the instabilities for the simplest case as shown in the dotted line 
box in Fig. \ref{Fig3schm_dimple2}b ($\theta=0$ and $\varphi=0$), and then 
generalize our results to arbitrary $\theta$ and $\varphi$.

\subsection{Domain Conversion and Step Pairing} \label{sec:domainconversion}

Let us begin with the simplest case, where the
vicinal surface is misoriented in the $\left[ 110\right] $ direction. At
equilibrium rather straight $S_{A}$ steps that run parallel to the dimer
rows of the upper $A$ terrace alternate with much rougher $S_{B}$ steps that
run perpendicular to the dimer rows of the upper $B$ terrace. \cite
{Si001_notation} When the field is normal to the steps, as illustrated in
the boxed region of Fig. \ref{Fig3schm_dimple2}b,
the terrace diffusion rates normal to the steps satisfy
$D_{t}^{B}\gg D_{t}^{A}$. We assume that the dimerization persists at
least to some extent on both adjacent half-step regions around each
terrace and will similarly affect diffusion rates there. The normal
diffusion in the two half-step regions around a given step is characterized by
$D_{s}^{A}$ and $D_{s}^{B}$. Taking account of the differences in
terrace diffusion rates, it seems reasonable to assume at least that
$D_{s}^{B} \geq D_{s}^{A}$, or
\begin{equation}
\left( D_{t}^{B}-D_{t}^{A}\right) \left( D_{s}^{B}-D_{s}^{A}\right)
\geq 0.  \label{eq:D_ineq}
\end{equation}
Special cases of this assumption include classical local equilibrium steps
where $R^{A}=R^{B}=1$ and a symmetric step model where $D_{s}^{B}=D_{s}^{A}$. 

The assumption here essentially states that the fundamental physics on Si(001) 
surfaces is dominated by the alternating reconstruction domains on terraces. 
Under this assumption, it is natural to think of the surface as made up of 
alternating $A$ and $B$ units, where the unit $\alpha$ ($\alpha=A$ or $B$) contains 
an $\alpha$ terrace together with the two neighboring $\alpha$ half step regions.   

We consider here cases where the system is driven away from equilibrium only
by the electric field. We make the usual assumption that permeability does
not play an essential role in this case and take the limit
$p_k \rightarrow 1$ for simplicity. We also
neglect evaporation and assume a positive effective charge.
The perfect sink limit decouples the concentration fields on the terraces, 
and permits a simple solution to the steady state diffusion problem in terms 
of exponential functions $e^{fx}$, where $f\equiv {\bf F}\cdot \hat{x}/k_{B}T$.

In almost all cases of experimental interest, the field is
sufficiently weak that $fs\ll fl_{t}\ll 1$, where $s$ and $l_{t}$ are the 
width of the step and terrace regions, and we obtain piecewise
linear profiles for the adatom concentration.
It is then straightforward to  write down the general solution for the
adatom density in unit $\alpha$ as
\begin{equation}
c^{\alpha }\left( x\right) =\left\{ 
\begin{array}{lc}
c_{eq}+m_{s}^{\alpha }\left( x+\frac{l_{t}^{\alpha }+s}{2}\right) &
-\frac{l_{t}^{\alpha } +s}{2}\leq x\leq -%
\frac{l_{t}^{\alpha }}{2} \\ 
c_{eq}+m_{t}^{\alpha }x & -\frac{l_{t}^{\alpha }}{2}\leq x\leq 
\frac{l_{t}^{\alpha }}{2} \\ 
c_{eq}+m_{s}^{\alpha }\left( x-\frac{l_{t}^{\alpha }+s}{2}\right) 
& \frac{l_{t}^{\alpha }}{2}\leq x\leq 
\frac{l_{t}^{\alpha }+s}{2}
\end{array}
\right.   \label{eq:c_linr}
\end{equation}
where $l_{t}^{\alpha }$ is the $\alpha$ terrace width. In the above
expression the origin is set at the center of the terrace region to take
maximum advantage of symmetry. It is easy to transform the origin
to the left atomic step position in accordance with the 
previous discussion on hopping models, and the results below will not be 
altered by any specific choice of the coordinate system. 

The $m_{s,t}^{\alpha }$ can be 
obtained by requiring continuity of concentration and flux 
at $\pm l_{t}^{\alpha }/2$ and are given by
\begin{eqnarray}
m_{t}^{\alpha }(l_{t}^{\alpha }) &=&c_{eq}sf\left( R^{\alpha
}-1\right) /\left( l_{t}^{\alpha }+R^{\alpha }s\right)  \nonumber  \\
m_{s}^{\alpha }(l_{t}^{\alpha }) &=&c_{eq}l_{t}^{\alpha }f\left(
1-R^{\alpha }\right) /\left( l_{t}^{\alpha }+R^{\alpha }s\right) 
\label{eq:slp_linr}
\end{eqnarray}
Here $R^{\alpha }\equiv D_{t}^{\alpha }/D_{s}^{\alpha }$
gives a dimensionless measure of the relative
diffusion rates in the $\alpha$ unit between the terrace and step regions in
a direction perpendicular to the step direction, and $c_{eq}$ is the average 
concentration for a uniform step array when $f=0$.

In the quasi-static approximation the step velocities are computed by a
flux balance. The surface flux normal to 
the step is constant throughout the $\alpha$ unit and is exactly given by 
\begin{equation}
J_{0}^{\alpha }=D_{t}^{\alpha }c_{eq}f\frac{ l_{t}^{\alpha }+s}{ l_{t}^{\alpha }+R^{\alpha }s}.
 \label{eq:J0}
\end{equation}
Because of the perfect sink assumption,
the fluxes in the individual $\alpha$ units on either side of a step are 
independent of each other. Thus the step velocity is easy to compute for a given
step configuration. 

Consider in particular the initial velocity of step $S_{\alpha}$ in a uniform
step train ($l_{t}^{A}=l_{t}^{B}=l_{t}$). This is given by
\begin{widetext}
\begin{eqnarray}
v_{0}^{\alpha} &=&\Omega \left( J_{0}^{\alpha}-J_{0}^{\beta}\right)  \nonumber \\
&=&\Omega c_{eq}f\left( l_{t}+s\right)  
{\displaystyle {\left[ \left( D_{t}^{\alpha}-D_{t}^{\beta}\right) l_{t}+\left( D_{s}^{\alpha}-D_{s}^{\beta}\right) sR^{\alpha}R^{\beta}\right]  \over \left( l_{t}+R^{\alpha}s\right) \left( l_{t}+R^{\beta}s\right) }}%
\label{eq:v0}
\end{eqnarray}
\end{widetext}
where $\alpha,\beta =A$ or $B$ and $\Omega$ is the atomic area. 
In this case the velocities of the two types of steps
satisfy $v_{0}^{B}=-v_{0}^{A}$. Therefore the initial uniform step array 
is not a steady state. Depending on the direction of the electric field,
one reconstruction domain expands while the other shrinks, creating step pairs
separated by the minor terrace. With a downhill current one finds
double height $D_{B}$ steps (consisting of an upper $S_{B}$ step and a lower 
$S_{A}$ step with a narrow $A$ terrace trapped in between) separated by
wide $B$ terraces; the equivalent configuration with $D_{A}$ steps and narrow
$B$ terraces is seen for an uphill current. Experiments show that
this field-driven step pairing continues until it is balanced at short
distances, probably by step repulsions, as first suggested by Natori {\em et
al.}, \cite{Si001pairingthy_Natori} in the special case where local equilibrium
was assumed for all the steps, corresponding to $R^{A}=R^{B}=1$ in our
model.

\subsection{Continued Bunching of Paired Steps} \label{sec:simulbunching}

\begin{figure}[tbp]
\includegraphics[width=76mm,height=66mm]{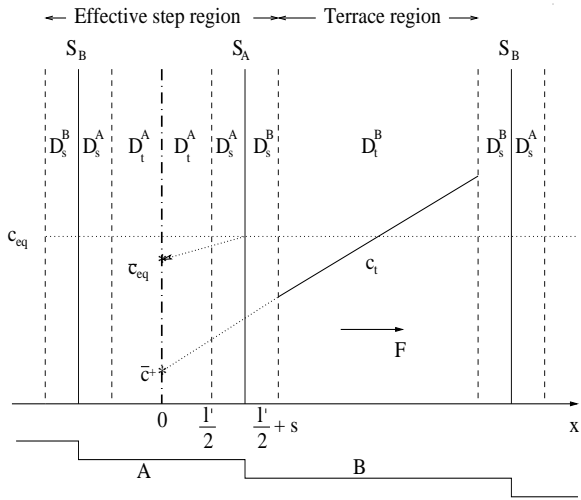}
\caption{A schematic illustration of extrapolation for an effective step
region. With a step-down current, domain $\left( 1\times 2\right) $ expands
to form an effective terrace region with some typical concentration profile
$c_{t}$. On the other hand, domain $\left( 2\times 1\right) $ shrinks
to $l^{\prime }$ and forms an effective step region when combined with the two
step regions bounding it. $c_{t}$ is extrapolated to the dotted-dashed line
at $x=0$ in the middle of the minor terrace which represents
the effective ``sharp" step.}
\label{Fig4schm_effregion}
\end{figure}

Now let us examine the stability of arrays of such paired steps.
Assuming that the step pairs (boundaries
of the minor domain) with constant spacings persist throughout the 
bunching process, as is shown by experiments, 
we can define a symmetric effective two-region model 
that can describe the continued bunching of
the paired steps. To that end, we treat the minor
reconstruction terrace together with the two step regions bounding it as
an effective step region that separates one major terrace from another,
as schematically shown in Fig. \ref{Fig4schm_effregion} for the case
of a step-down current. As shown below (and previously discussed \cite{TZ_em_short}), 
the bunching behavior is determined by the field direction and the sign
of the kinetic coefficient for the sharp step 
model associated with the effective two-region model defined here.

In the Appendix we discuss the mapping to a sharp step model from a 1D
hopping model that treats the effects of the electric field explicitly,
while assuming there is a perfect sink at $x=0$ in the center of the
step region. In the present case a minor terrace resides at the center
of the effective step region. We can still follow basic treatment in the
Appendix if we take this into account by
shifting the origin of the coordinate system by transforming
$x\rightarrow x-\left( l^{\prime }+s\right) /2$ and $s/2\rightarrow
l^{\prime }/2+s$, where $l^{\prime }$ is the width of the minor
reconstruction domain. Representing the discrete terrace concentrations
by a continuum function $c\left( x\right) $ and Taylor expanding as
before about $x=0^+$ --- the center of the effective step region ---
we find that the
effective sharp interface boundary condition takes a form analogous to
Eq. (\ref{eq:BCFemps}):
\begin{equation}
D_{t}^{\alpha }\left[ \nabla c\lvert_{+}- f\overline{c}^{+}\right] =
\bar{k}^{\alpha }\left( \overline{c}^{+}-\bar{c}_{eq}\right) ,  \label{eq:bc_eff}
\end{equation}
where 
$D_{t}^{\alpha }$, $\alpha=A$ or $B$, is
the normal diffusion constant in the major
terrace. The continuum profile is schematically depicted in Fig. \ref
{Fig4schm_effregion} for the case of a step-down current.  

Two new features are seen in Eq. (\ref{eq:bc_eff}) arising from the use of a
single effective step region to describe the paired steps.
First, the major terrace is determined by the
current direction in the initial step pairing
regime.  Second, both an effective kinetic coefficient $\bar{k}$ and an
effective ``equilibrium concentration'' $\bar{c}_{eq}$ appear in the sharp
step boundary condition. The latter is given by
\begin{equation}
\bar{c}_{eq}=c_{eq}\left[ 1-\frac{1}{2}f\left( l^{\prime }+s\right) \right] .
\label{eq:ceq_eff}
\end{equation}
This can be understood heuristically by noting that the
effective density in the center $\bar{c}_{eq}$ 
should be linearly modified by the weak field from its value $c_{eq}$
at the ``real'' local 
equilibrium step near the lower boundary of the effective step region.
Similarly, the effective attachment length 
$\bar{d}^{\alpha }$ associated with the effective kinetic
coefficient $\bar{k}^{\alpha }$
is given by 
\begin{equation}
\bar{d}^{\alpha }\equiv
\frac{D_{t}^{\alpha }}{\bar{k}^{\alpha }}=\frac{\bar{s}}{2}\left[ \bar{R}^{\alpha }-1
\right] ,  \label{eq:map_eff}
\end{equation}
where $\bar{s}=l^{\prime}+2s$ is the width of the effective step region and 
$\bar{R}^{\alpha }=sR^{\alpha }/\bar{s}$ is the relative diffusivity in the 
effective two region model defined above. 

Equations (\ref{eq:bc_eff})-(\ref{eq:map_eff}) give the mapping between 
the effective two region model describing paired steps separated by major
terraces and an equivalent sharp step model. In the 
steady state where the major terraces all have the same width,
the surface flux in the sharp step model can be
obtained from Eq. (\ref{eq:J0}) as follows. We replace the 
parameters $c_{eq}$, $s$, and $R^{\alpha }$
by the corresponding effective parameters $\bar{c}_{eq}$, $\bar{s}$, and 
$\bar{R}^{\alpha }$. Clearly $l=l_{t}+\bar{s}$ represents the terrace width 
in the sharp step model.
The steady state flux in the sharp 
step model as a function of the terrace width is thus given by 
\begin{equation}
J_{0}^{\alpha }(l)=D_{t}^{\alpha }c_{eq}f\frac{l}{l+
2\bar{d}^{\alpha }}. 
\label{eq:Jeff}
\end{equation}
Note that $\alpha =A$ or $B$ is determined by the current direction.

To examine the stability of the above steady state, consider a small 
deviation $\delta x_{n}=\varepsilon_{n}e^{\omega t}$ of the $n^{th}$ step 
in the uniform step train, where $\varepsilon_{n}=\varepsilon e^{in\phi}$. 
Here $\varepsilon$ is a small constant
and $\phi$ is the phase between neighboring 
steps. Then the $n^{th}$ step will move in response to the unbalanced flux 
induced by the changed widths of the terraces in front $l_{n}=l+\varepsilon_{n}
(e^{i\phi}-1)$ and back $l_{n-1}=l+\varepsilon_{n}(1-e^{-i\phi})$. The linear 
amplification rate $\omega=v_{n}/\varepsilon_{n}$ is given by 
\begin{equation}
\omega =\Omega D_{t}^{\alpha }c_{eq}\frac{4\bar{d}^{\alpha }f}{\left( l+2\bar{d}^{\alpha }
\right) ^{2}}\left( 1-\cos \phi \right).  \label{eq:bch_bcf}
\end{equation}
An instability towards step bunching results if
$\bar{d}^{\alpha }f>0$ with a maximum
at $\phi=\pi$, corresponding to step pairing. Note that the direction of
the field and the sign of the effective kinetic coefficient combine
to determine when step bunching occurs, as discussed earlier. \cite {TZ_em_short}
       
Using Eqs. (\ref{eq:map_eff}) and (\ref{eq:bch_bcf}),
we see that to get simultaneous step
bunching from current in {\em both} directions, as seen in experiment, requires 
\begin{equation}
R^{A}>2+\frac{l^{\prime}}{s}>R^{B}.  \label{eq:R_crit}
\end{equation}
With a step-down current, the first part of the inequality in Eq. (\ref
{eq:R_crit}) makes the effective kinetic coefficient for the effective step
region containing the slower diffusion domain positive, which results in
a step bunching instability. The second inequality in Eq. (\ref
{eq:R_crit}) give rise to a {\em negative} effective kinetic coefficient which
produces step bunching with a step-up current.
Note that this does not require negative kinetic coefficients 
for single steps of either kind. 

However, if one assumes the individual steps are at local equilibrium,
($R^{A}=R^{B}=1$), then the kinetic coefficient for the effective step region
is negative in both cases, and therefore bunching is expected only from a
step-up current.

\subsection{Angular Dependence}

It is straightforward to extend the above analysis to a general dimple
geometry, where the domain conversion exhibits interesting angular
dependences. A schematic view of the experimental dimple is shown in Fig. \ref
{Fig3schm_dimple2}a. Again, we need to consider the fluxes from the neighboring
terraces going into the step. Using Eq. (\ref{eq:J0}), we can represent
the surface flux as the sum of fluxes along the two characteristic
directions, 
\begin{equation}
{\bf J}_{f}=\cos \theta J_{0}^{B}\widehat{\parallel}+\sin \theta
J_{0}^{A}\widehat{\perp}  \label{Jf}
\end{equation}
for the front terrace and 
\begin{equation}
{\bf J}_{b}=\cos \theta J_{0}^{A}\widehat{\perp}+\sin \theta
J_{0}^{B}\widehat{\parallel} \label{Jb}
\end{equation}
for the back terrace of step $1$ in Fig. \ref{Fig3schm_dimple2}a, where $\parallel$ 
and $\perp$ are the directions parallel and perpendicular to dimer rows as 
defined earlier.
The angular dependent step velocity is readily obtained 
\begin{equation}
v_{0}^{(1)}\left( \theta ,\varphi \right) =v_{0}^{B}\cos \left( 2\theta -
\varphi \right) ,  \label{eq:v_thetaphi}
\end{equation}
where $v_{0}^{B}$ is given by Eq. (\ref{eq:v0}). Eq. (\ref{eq:v_thetaphi}) 
shows that a steady state of paired steps will 
form on the part of the dimple where $\cos \left( 2\theta -\varphi\right)\neq 0$.
  
In the following, we will concentrate on two special configurations that are 
studied experimentally. \cite{Si001bchexp_Nielsen}  
The first is shown in Fig. \ref{Fig3schm_dimple2}b, where the current is 
parallel to the dimer row direction. In this case $\theta =0$, and
$\cos \varphi $ characterizes the angular 
dependence around the dimple. The maximum pairing instability occurs at 
$\varphi =0$ where the current is perpendicular to the step normal direction, 
and no instability in seen at $\varphi =\pm \pi/2$. From the previous 
discussion in Sec \ref{sec:domainconversion} and Sec \ref{sec:simulbunching}, 
we can easily see that continued step bunching occurs with a 
maximum at $\varphi=0$. 

The other interesting configuration corresponds to an upright field parallel 
to $\left[ 010 \right]$ direction in Fig. \ref{Fig3schm_dimple2}a. 
In this case the current is at an angle $\theta =\pi/4$ from the 
dimer row direction. Hence the angular dependence becomes 
$\cos \left( \pi /2-\varphi \right) =\sin \varphi $. The maximum pairing 
instability occurs at $\varphi =\pi /2$, where the current is {\em parallel}
to the steps, and no instability occurs when the current is
perpendicular to the steps. Again the sharp step model corresponding to 
the steady state can be extracted.  The subsequent step bunching 
instability for a parallel current was discussed by Liu \textit{et al.}
\cite{Stepbendingthy_Liu} Their stability analysis suggests that step 
bunching generally occurs for a non-vanishing attachment/detachment 
length $d$, regardless of its sign, when the current is parallel to the 
average step positions.

The results discussed here are in good agreement with experiments. For the 
angular dependent step pairing, the result is consistent with the original 
analysis by Nielsen \textit{et al.}.\cite{Si001bchexp_Nielsen} 
However, our explanation for the subsequent step bunching is different. 
Our analysis provide a simpler scenario that does not require a tensor 
character to the effective charge.

\section{Conclusion}

This paper derives expressions for sharp step boundary conditions characterized by 
linear kinetics rate parameters $k_{\pm}$ and ${P}$ for general BCF 
type models by appropriate coarse-graining from a microscopic hopping model. 
$k_{\pm}$ and ${P}$ are related to the attachment/detachment kinetics at kinks and 
to diffusion across the ledges respectively. In particular, our study shows  
that both parameters can be negative when diffusion is faster in the step
region than on terraces.  The possibility of negative kinetic coefficients
was first suggested by Politi and Villain, \cite{negativecoef}
but with no derivation or discussion 
of any physical consequences. In the appropriate limit, we recover 
the mapping previously obtained with the CTRM.\cite{TZ_em_short}
Our results also seem consistent 
with those from phase field models, \cite{Phasefieldsteps_Pierre} 
while providing a simple and physically suggestive picture.

We then used the perfect sink limit ($p_{k}=1$) of the general model to analyze 
current-induced instabilities on Si(001),
where the field represents the only driving force away from equilibrium,
and found results in good agreement 
with experiment. As we will show elsewhere, this same limit also provides a coherent
interpretation of Si(111) electromigration experiments,
where novel step wandering behavior is seen. \cite{em_long} 
Thus we believe that the perfect sink model provides a simple and physically 
reasonable description of many electromigration experiments on vicinal Si
surfaces.

\begin{acknowledgments}
We are grateful to Ted Einstein, Oliver Pierre-Louis, and Ellen Williams 
for stimulating discussions. This work has been supported by the NSF-MRSEC
at the University of Maryland under Grant No. DMR 00-80008.
\end{acknowledgments}
\appendix


\section{1D Perfect Sink Model with a Constant Electric Field}

In the generic hopping model discussed earlier, we assumed that the flux arose
only from concentration gradients. We consider here the case where there
is a additional external driving force from the electric field,
and take the perfect sink limit $p_k=1$ used to analyze experiments.
In particular, we examine whether or not the kinetic rate parameters in 
the resulting sharp step model could depend on the field as Suga \textit{et al.} 
previously suggested.\cite{Si111bchthy_Suga}

In the absence of the field, the 1D potential energy surface is similar to that
in Fig. \ref{Fig1nonps_generic}, where now the site at $x=0$ is a perfect sink
surrounded by a more general region of width $s$ with
different diffusion barriers.
When a weak electric field is applied in the positive $x$ direction, 
the potential
energy surface will be modified by an amount $V=-\int {\bf F}d{\bf x}$,
where ${\bf F}=z^{*}e{\bf E}$, $z^{*}e$ is the effective charge. 
The modification of the potential surface produces a bias for adatom 
hopping, which will later lead to a convective flux contribution in the 
continuum description.

The driven flux inside the step region can be written as 
\begin{equation}
J_{x+a/2}=\frac{D_{s}}{a}e^{fa/2}\hat{c}_{s}\left( x\right) -\frac{D_{s}}{a}
e^{-fa/2}\hat{c}_{s}\left( x+a\right) ,  \label{eq:Jx+a/2}
\end{equation}
where $f\equiv \left| {\bf F}\right| /k_{B}T$. The quasi-static
approximation suggests continuity of fluxes, i.e. $J_{x+a/2}=J_{x-a/2}$,
which leads to the following equation for the discrete concentration 
$\hat{c}_{s}\left(x\right) $,
\begin{widetext}
\begin{equation}
0 =e^{-fa/2}\hat{c}_{s}\left( x+a\right) 
-\left( e^{fa/2}+e^{-fa/2}\right) \hat{c}_{s}\left( x\right)
+e^{fa/2}\hat{c}_{s}\left( x-a\right) ,
\label{eq:cseqn}
\end{equation}
\end{widetext}
where $x$ is evaluated at discrete lattice sites inside the step region. It is
easy to write down the solution of Eq. (\ref{eq:cseqn}) as 
\begin{equation}
\hat{c}_{s}\left( x\right) =c_{eq}+A\left( e^{fx}-1\right) ,
\label{eq:cs_sln_gen}
\end{equation}
taking account of the perfect sink at $x=0$.
Here $A$ is a constant that can be determined
by continuity of fluxes at the boundary between step and terrace region,
i.e. $J_{s/2-a/2}=J_{s/2+a/2}$. This gives 
\begin{equation}
A=\frac{e^{-fa/2}R\hat{c}_{t}\left( \frac{s}{2}+a\right) -\left[
e^{-fa/2}+e^{fa/2}\left( R-1\right) \right] c_{eq}}{e^{fa/2}\left(
e^{fs/2}-1\right) R+e^{fa/2}-e^{-fa/2}}.  \label{eq:A}
\end{equation}
Here $\hat{c}_{t}$ is the discrete concentration on the terrace site.

To obtain the sharp interface boundary condition, we apply flux
continuity $J_{s/2+a/2}=J_{s/2+3a/2}$, and express all the discrete 
terrace concentrations in terms of the extrapolated $c^{+}$ and the 
corresponding gradient $\nabla c\mid _{+}$. In 
the weak field limit that is valid in most experiments,
we can linearize the exponentials in all of the above 
expressions. To the leading order, we obtain the boundary condition as 
\begin{equation}
\pm D_{t}\left[ \nabla c\mid _{\pm }\mp fc^{\pm }\right] =k_{\pm }\left( c^{%
\pm }-c_{eq}\right).  \label{eq:BCFemps}
\end{equation}
where results for both the $+$ and $-$ sides can be given by symmetry.
Note that the term proportional to $f$ is the convective flux 
induced by the field, which is of the same order as the concentration 
gradient. As mentioned earlier in Section \ref{sec:discretecont},
the mapping to the kinetic coefficient is independent 
of the field to lowest order, and is given by  
\begin{equation}
d_{\pm }\equiv \frac{D_{t}}{k_{\pm }}=\frac{1}{2}\left( R_{\pm }-1\right) s,
\label{eq:emmapping}
\end{equation}
where $R_{\pm }\equiv D_{t}/D_{s}$. Equation (\ref{eq:emmapping}) recovers
the results we derived earlier from CTRM, and is also consistent with the general
result in Eq. (\ref{dps}). 


\end{document}